\documentclass[aps,twocolumn,groupedaddress]{revtex4}
\usepackage{graphicx}

\begin{document}

\title{Electron-magnon coupling and 
nonlinear tunneling transport in magnetic nanoparticles}


\author{\L.~Michalak}
\author{C.~M.~Canali}
\affiliation{Department of Chemistry and Biomedical Sciences, Kalmar
University, 31912 Kalmar, Sweden}
\author{V.G.~Benza}
\affiliation{Dipartimento di Fisica e Matematica, Universita' dell'Insubria,
Como, Italy}



\begin{abstract}
We present a theory of single-electron tunneling transport through 
a ferromagnetic nanoparticle in which particle-hole excitations are coupled
to spin collective modes. 
The model employed to describe the interaction between quasiparticles and
collective excitations captures the salient features of a recent microscopic
study. 
Our analysis of nonlinear quantum transport 
in the regime of weak coupling to the external electrodes 
is based on a rate-equation formalism for the nonequilibrium 
occupation probability of the nanoparticle many-body states. 
For strong electron-boson coupling, 
we find that the tunneling conductance 
as a function of bias voltage is characterized by a large and dense
set of resonances. Their magnetic field dependence in the large-field
regime is linear, with slopes of the same sign.  Both features are in 
agreement with recent tunneling experiments.

\end{abstract}

\maketitle

\vskip2pc

Metallic nanoparticles are among the best physical 
realizations of the concept of Fermi liquid introduced by Landau more than
fifty years ago. Their discrete low-energy spectra can be put in a 
one-to-one correspondence with those of corresponding noninteracting electron
systems. 
Single-electron tunneling spectroscopy\cite{cbreviews}
in normal-metal nanograins provides a vivid example of
Landau's enormous simplification 
of interacting Fermi systems.
Most of the interesting phenomena studied in these experiments can indeed be 
understood in terms of the quantum mechanics of confined
noninteracting quasiparticles.
If the grain is made of a ferromagnetic transition-metal material, 
however, the discrete resonant spectrum seen in 
tunneling experiments\cite{cornell1,cornell2}
is far more complex than the one predicted in an
independent particle picture, and 
indicates that the quasiparticle states are coupled to the 
collective magnetic moment of the grain.
Since ferromagnetic transition metals,
in addition to Landau's particle-hole (p-h) excitations, 
support low-energy collective spin excitations, it is reasonable to assume
that tunneling transport through ferromagnetic nanoparticles involves
some kind of spin excitations
that are the finite-system analogue of the familiar
spin-waves or magnons of bulk ferromagnets. 
So far attempts of 
including spin collective modes in tunneling transport based on a 
simple toy model\cite{cmc_ahm2000prl,kleff2001prb,kleff_vdelft2001prb} 
have 
explained only in part the rich phenomena seen in experiment.

In this paper we present a theoretical study of single-electron 
tunneling transport through a ferromagnetic metal nanoparticle
based on a model that captures the salient features of
its elementary excitations -- p-h and spin collective --
as derived from a recent {\it microscopic} study\cite{cmc_ac_ahm2002pap4}.
A few remarkable features seen in experiment
emerge in a very transparent and direct way from our treatment of the
electron-magnon coupling.
We find that when a low-energy p-h excitation is
strongly coupled to one of the spin collective modes, the tunneling
differential conductance versus bias voltage displays an enhanced density
of resonances with spacings smaller than
 the independent-electron
energy mean-level spacing $\delta$. 
The dependence of the tunneling resonances 
on external magnetic field is regulated by the behavior of the
underlying quasiparticle states; it is characterized by mesoscopic 
fluctuations
at small fields and a monotonic dependence at fields larger than the
switching field. 
The model further predicts that in the limit of ultrasmall nanoparticles,
where $\delta$ is much larger than the typical
magnon energy, the conductance should display clusters of resonances 
separated by an energy of order $\delta$.


The choice of our model is motivated by the microscopic analysis of
Ref.~\cite{cmc_ac_ahm2002pap4}, where the explicit derivation of the
exchange-field-fluctuation propagator allows one to determine the 
elementary spin excitations (Stoner p-h and collective) of a magnetic 
grain.
One finds that for a small nanoparticle there is one isolated 
spin collective mode
below the lowest p-h excitation energy, which 
corresponds to the
ferromagnetic resonance excitation (spatially uniform $q=0$ spin-wave),
of energy $E_{\rm res} \sim $ magnetic anisotropy energy/atom 
$\approx 0.1$ meV in cobalt.
For large nanoparticles, the ferromagnetic resonance lies in a region of 
p-h quasicontinuum and acquires a line-width 
$\alpha E_{\rm res}$, where $\alpha <1$ is 
the Gilbert damping parameter. 
The crossover between these two regimes occurs
when one p-h excitation contributes to the
resonance, namely when $\delta = \sqrt{\alpha} E_{\rm res}$.
Although the nanoparticles investigated in  
Ref.~\cite{cornell1, cornell2} are too small
to strictly satisfy this condition, 
interactions between p-h excitations
and spin wave modes, including the {\it nonuniform}
ones ($ q \ne 0$), will frequently occur.  
As long as the mechanism of the interaction of one p-h
excitation with one spin-wave mode is independent on 
the latter being uniform or nonuniform, we can illustrate it 
by following 
Ref.~\cite{cmc_ac_ahm2002pap4}, where
the uniform case was considered.
It was shown that when only one p-h excitation of energy 
$\epsilon_{ab}= \xi_b - \xi_a$
is close to $E_{\rm res}$,
the exchange-field propagator has two poles at energies
\begin{equation}
\omega_{\pm} = \frac{E_{\rm res}+\epsilon_{ab}}{2} \pm
\big[ [(E_{\rm res}-\epsilon_{ab})/2]^2 + \gamma^2 \big]^{1/2}\;.
\label{energies}
\end{equation}
The avoided crossing gap $\gamma$ resulting from the collective mode 
p-h coupling is found to be 
$\gamma \sim \alpha E_{\rm res}= 
{\Delta_{\rm MF}\over \sqrt{2S}}|\langle b | S_x | a \rangle |$,
where $2S = N_{\uparrow} - N_{\downarrow}$ is the 
total spin of the nanoparticle,
and $\Delta_{\rm MF}$ is the amplitude of the spin-splitting
field.
The presence of the matrix element $|\langle b | S_x | a \rangle |$
in the expression for $\gamma$ emphasizes the fact that the 
coupling between spin-waves or magnon and electrons
is ultimately due to the exchange interaction, which conserves spin. 
Thus the quasiparticle states $a$ and $b$ should 
have opposite spins. 
In spite of the fact that most of the states lying close to the
Fermi level have minority-spin character, because of spin-orbit coupling, 
the quasiparticle states $\vert I \rangle$ 
are in fact linear combinations of spin-up and spin-down components
$\vert I \rangle = 
\alpha_I \vert \uparrow \rangle + \beta_I \vert \downarrow  \rangle$,
and the matrix element $|\langle b | S_x | a \rangle |$ will not vanish. 
Notice, however, that since spin-orbit interaction is relatively 
weak in transition-metal 
ferromagnets, the quasiparticle states can still be assumed 
to have in general one predominant spin
character with just a small admixture of the other.  


 
\begin{figure}
 \includegraphics[width=3.0in,height=1.5in]{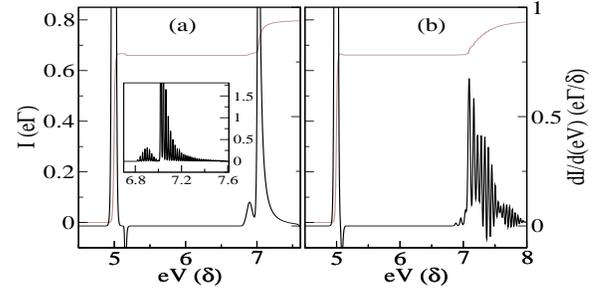}
 \caption{Current and differential conductance versus bias voltage $V$
for $\omega = 0.1 \delta$.
(a) The electron-magnon coupling strength is $\gamma= 0.1\delta$;
(b) $\gamma= 0.2 \delta$.
The temperature $T$ is set equal $0.005\,\delta$, except in the inset, where
it is equal to $ 0.001\,\delta$.} 
 \label{fig1}
 \end{figure}

The Hamiltonian describing the 
isolated nanoparticle in which 
a magnon
is coupled to one
p-h excitation is
\begin{equation}
H_d = \sum_{i=a,b}\epsilon_i c^\dag _i c^{\phantom{\dagger}}_i  + 
 \omega \beta^\dag \beta^{\phantom{\dag}} + 
\gamma (c^\dag_a c^{\phantom{\dagger}}_b \beta^\dag + 
c^\dag_b c^{\phantom{\dagger}}_a \beta) 
+ U \hat{n} (\hat{n} - 1),
\label{dotham}
\end{equation}
where $c^\dag _i$ and $c^{\phantom{\dagger}} _i$ with $i=a,b$ are Fermi 
operators creating and annihilating two electronic levels of energy
$\epsilon_a$ and $\epsilon_b$ respectively, with $\epsilon_a <\epsilon_b$. 
The Bose operators $\beta^\dag$ and
$\beta$ describe a magnon 
of energy $\omega$. Below we measure all energies in units of
the mean-level spacing $\delta \equiv \epsilon_b-\epsilon_a$.
The term 
$\gamma\, (c^\dag_a c^{\phantom{\dagger}}_b \beta^\dag + 
c^\dag_b c^{\phantom{\dagger}}_a \beta)$
represents
the electron-magnon coupling. 
It can be interpreted as a vertex describing
an electron scattering from the electronic state $a$ (respectively, $b$) to the 
state $b$ ($a$), 
while absorbing (emitting)
a magnon. 
We will view the coupling strength $\gamma$ as a phenomenological
parameter; $\gamma \sim \omega$ represents strong coupling.
Recently electron-boson interactions have been
used extensively to model electron-phonon coupling  
in molecular single-electron transistors\cite{el_phonon}. An 
interaction term more similar to ours has been used in studying
magnon-assisted transport
in ferromagnetic tunneling junctions\cite{falko}.
The last term in Eq.~\ref{dotham} 
represents a Coulomb repulsion energy, which is nonzero when both
electronic levels are occupied, $\langle \hat n\rangle=2$. 
The model in Eq.~\ref{dotham}, 
representing a double-level system coupled to one boson mode, is well known
in quantum optics and cavity quantum electrodynamics under the name of
Jaynes-Cummings model \cite{JC63}.
The model can be solved exactly, since it conserves both the
number of electrons $n = n_a + n_b $ and the quantity $ (n_b - n_a)/2 + m$,
where $m$ is the number of bosons.
In the trivial cases  $n=0$ and $n=2$
the energy spectrum is 
$\epsilon ^{n}_m =  \omega m + n/2(\epsilon_a + \epsilon_b + 2U)$;
the 
corresponding eigenstates are 
$\vert 0, m\rangle = 
(\beta^{\dagger})^m\vert 0\rangle$
and $\vert 2, m\rangle = 
c^{\dagger}_ac^{\dagger}_b(\beta^{\dagger})^m\vert 0\rangle$, where
$\vert 0\rangle$ is the vacuum.
The $n=1,(n_b-n_a)/2 + m +1/2 = k+1$ eigenspace 
is spanned by the states: 
\begin{equation}
\vert 1_a, k+1\rangle \equiv 
c_a^{\dagger}\; (\beta^{\dagger})^{k+1}\;\vert 0\rangle\;,\ \ \ 
\vert 1_b, k\rangle \equiv 
c_b^{\dagger}\; (\beta^{\dagger})^k\;\vert 0\rangle\;.
\end{equation}
The Hamiltonian is now diagonalized
within each $k$-subspace, yielding the eigenvalues 
$\epsilon_k^+$ and $\epsilon_k^-$
\begin{equation}
\epsilon^{\pm}_k = \epsilon^0_k +
\epsilon_{\rm av} \pm \frac{1}{2} \sqrt{\epsilon_{\rm res}^2 + 
4 \gamma ^2 (k + 1)}\;,
\label{eingenenergies}
\end{equation}
where $ \epsilon_{\rm res}\equiv (\epsilon_b-\epsilon_a) - \omega $ 
and $\epsilon_{\rm av} \equiv \frac{1}{2} (\epsilon_a + \epsilon_b + \omega)$.
The corresponding eigenvectors are 
\begin{equation}
|\pm, k \rangle = \delta^{\pm}_1(k) |1_a, k + 1 \rangle +  \delta^{\pm}_2(k) |1_b, k \rangle\;,
\label{simpleig}
\end{equation}
where
\begin{eqnarray}
\delta^{\pm}_1(k) = 
{\gamma \sqrt{k+1}\over
{\sqrt{[\epsilon^{\pm}_k - \epsilon_a -  \omega 
(k + 1)]^2 + 
\gamma ^2 (k + 1)}}}
\;,\\
\delta^{\pm}_2(k)=
{[\epsilon^{\pm}_k - \epsilon_a - 
 \omega (k + 1)]\over
{\sqrt{[\epsilon^{\pm}_k - \epsilon_a -  \omega 
(k + 1)]^2 + 
\gamma ^2 (k + 1)}}}
\;.
\end{eqnarray}
On top of these states $\vert \pm, k\rangle$ there is also the state
$\vert 1_a, 0\rangle$ with energy $\epsilon_a$, which forms a decoupled
one-dimensional subspace in the $n=1$ sector.
We now assume that the magnetic grain is weakly coupled to metallic 
external electrodes and investigate single-electron tunneling transport
through the grain \cite{cbreviews}.
The total Hamiltonian describing the system is
$H = H_d  + H_l + H_r + H_t$,
where $H_d$ is given in Eq.~\ref{dotham};
$H_l$ and $H_r$ describe the left and right 
electrodes, assumed
to be normal Fermi liquids 
$H_{\alpha}=\sum_{p} 
\xi_{p\alpha} c^\dag _{p\alpha} c^{\phantom{\dagger}}_{p\alpha}$, 
$\alpha = l, r$,
where $p$ is the quantum number specifying a quasiparticle of energy  
$\xi_{p\alpha}$ measured with respect to the chemical potential 
of lead $\alpha$; 
$H_t$ is the tunneling Hamiltonian
coupling the grain to the electrodes
$H_t = \sum_{p,\alpha=r,l} \Big[t_{p\alpha} c^\dag _{p\alpha} 
(c^{\phantom{\dagger}}_a + c^{\phantom{\dagger}}_b) + {\rm h.c.}\Big]$.
In the limit of weak coupling, transport takes place via sequential tunneling,
which can be described by means of a standard
rate-equation formalism for the occupation
probabilities of the grain many-body states \cite{rate_equation}.
We are interested in the regime
where Coulomb blockade is first lifted by applying an external bias voltage,
and only the two charge states $n=0, 1$ are involved. The master equations
describing the kinetics of the {\it nonequilibrium} occupation probabilities 
$P_k^n = \{P^0_k\,, P^a_0\,, P^-_k\,, P^+_k\}$ for the states 
$\{\vert 0, k\rangle, \vert 1_a, 0\rangle, 
\vert -, k\rangle,
\vert +, k\rangle,\ k= 0,1,\dots\}$, are
\begin{eqnarray}
\dot{P^0 _k} &=& - \sum_{k';\alpha} \Big[2P^0 _k (W^{\alpha}_{0k;+k'}
+ W^{\alpha}_{0k;-k'} + W^{\alpha}_{0;a})\nonumber\\ 
\label{master1}
&&\ \ \ \ \ \  - P^+ _{k'} 
W^{\alpha}_{+k';0k} - P^- _{k'} W^{\alpha}_{-k';0k} - P^{a}_0 W^{\alpha}_{a;0} \Big],\\
\dot{P^{a} _0} &=&  = 2 P^0 _0 \sum_{\alpha} W^{\alpha}_{0;a} - P^{a}_0 \sum_{\alpha} W^{\alpha}_{a;0},\\
\label{master2}
\dot{P^{\pm} _k}&=& - \sum_{k';\alpha}\Big[P^{\pm} _k W^{\alpha}_{{\pm}k;0k'} - 
2P^0 _{k'} W^{\alpha}_{0k';{\pm}k}\Big].
\label{master3}
\end{eqnarray}
The coefficients $W^{\alpha}_{\dots}$ appearing in Eqs.~\ref{master1}-\ref{master2}
are transition rates between two many-body states of the grain
caused by electron tunneling from and to the leads. 
For instance, $W^{\alpha}_{0k;+k'}$ is the transition rate 
from state $|0, k\rangle$ to $|+, k'\rangle$ 
due to an electron tunneling from the $\alpha$-electrode onto the grain. 
The $W^{\alpha}_{\dots}$ are given by Fermi's golden rule

\begin{eqnarray}
W^{\alpha}_{0k;\pm k'} &=& \Gamma^{\alpha} _{0k;\pm k'}\; n_{\rm F} 
(\epsilon^{\pm}_ {k'} - \epsilon^0_k - \mu_{\alpha})\nonumber\\ 
&{\phantom = }&\times[\delta_1^{\pm}(k')^2\; \delta_{k',k-1} + \delta_2^{\pm}(k')^2\; \delta_{k',k}]
\; ,\\
\label{tunrates1}
 W^{\alpha}_{0;a} &=& \Gamma^{\alpha} _{0;a} n_{\rm F} 
(\epsilon_a -  \mu_{\alpha})\;,  
\label{tunrates3}
\end{eqnarray}
where $\mu_\alpha$ is the electrochemical potential of lead $\alpha$, which
we assume to be shifted symmetrically around zero by the applied bias voltage
$V$: $\mu_{\rm l} = - \mu_{\rm r} = V/2$. 
The transition rates $W^{\alpha}_{\pm k';0k}$ and $W^{\alpha}_{a;0}$ 
are obtained
from $W^{\alpha}_{0k;\pm k'}$ and $W^{\alpha}_{0;a}$ respectively,
by replacing
the Fermi function with 
$[1 - n_{\rm F}]$ evaluated at the same energy.
The tunneling rates
$\Gamma^{\alpha} _{0k;\pm k'} = \frac{2\pi}{\hbar} \sum_{p} |t _{p \alpha}|^2 \delta(\epsilon_{p \alpha} - (\epsilon^{\pm} _{k'} - \epsilon_k))$ will be
taken for simplicity to be independent of energy and lead index,
$\Gamma^{\alpha} _{0k;\pm k'} = \Gamma$.
The nonequilibrium steady-state
probability $P^n_k$ is the solution of the
matrix equation $\breve{M} \breve{P} = 0$, 
where the matrix $\breve{M}$ includes all the transition rates, 
and $\breve{P}$ is a vector of all the $P^n_k$'s.
The dc current through the left or right junction is then written as
\begin{eqnarray}              
I& =& (+/-) e\; \Big\{\sum_{k}\Big [2 P^0 _k \sum_{k'} (W^{l/r} _{0k;-k'} 
+ W^{l/r} _{0k;+k'})\nonumber\\
&-& P^- _k \sum_{k'} W^{l/r} _{-k;0k'}
-P^+ _k \sum_{k'} W^{l/r} _{+k;0k'}\Big]\nonumber\\
&+& 2 P^0_0 W^{l/r} _{0;a} - P^a_0 W^{l/r} _{a;0}\Big\}\;.
\label{current}
\end{eqnarray}
The transition sequence  
$\vert n=0\rangle \rightarrow \vert n=1\rangle \rightarrow \vert n=0\rangle$
allows the tunneling electron to probe the
coupled p-h spin-wave excitations of the grain, which appear as resonances
in the differential
conductance $dI/dV$ as a function of the bias voltage $V$.
We discuss first the case where the p-h excitation is coupled with the uniform
($q=0$) spin-wave mode. For the nanoparticles considered in 
Refs.~\cite{cornell1, cornell2},
$\delta \sim $ 1 meV, 
while the energy of the uniform spin-wave is approximately equal to
the anisotropy energy/atom $\sim 0.1$ meV.  
In Fig.~\ref{fig1} we plot $I$ and $dI/dV$ vs $V$ for the case
$\omega = 0.1 \delta$, which pertains to this situation.
The calculations are done at temperature
$T= 0.005\delta$, corresponding to the experimental
$T\approx 50$mK.
When $\gamma = \omega$ [ Fig.\ref{fig1}(a) ], three sets of peaks 
in the conductance are visible.
The first isolated peak occurs when the current starts to flow,
and corresponds
to the successive transitions 
$\vert 0\rangle \rightarrow \vert 1_a ,0\rangle\rightarrow  \vert 0\rangle$
which are possible when $\mu_l = V/2 = \epsilon_a$\cite{foot1}. 
On further increasing $V$,
the current remains 
constant until the next lowest charging
state $ \epsilon^{-}_0$ becomes available (at $eV = 6.8 \delta$ for this case).
For yet larger $V$ 
higher states
$\vert 0 ,k\rangle$ and $\vert - ,k\rangle$
acquire a finite nonequilibrium occupation
probability, and new transport channels open up.  
In principle, each allowed transition  
$\vert 0 ,k\rangle \rightarrow \vert - ,k\rangle$ gives a resonance
at $ \epsilon^{-}_k - \epsilon^{0}_k$, as shown in the inset of 
Fig.~\ref{fig1}(a), calculated 
at very low temperature, $T= 0.001 \delta$. But at
$T = 0.005 \delta$ only their envelope is visible in the form 
of a small bump in the
conductance centered at $e {\rm V} = 6.9\delta$. The third large peak,
appearing at 
$e {\rm V} = 7.0\delta$ 
is also the envelope of many closely spaced resonances, 
caused primarily by the transitions through the second group of charged
states, $\vert + ,k\rangle$, which become available at that energy.
Although values of $\gamma > \omega $ are not very realistic, 
it is instructive to study the limit behavior of the tunneling conductance for
large values of the magnon-electron coupling. In Fig.~\ref{fig1}(b) we plot
$I$ and $dI/dV$ vs. $V$ for $\gamma= 2\omega$. We can see that a large $\gamma$
causes the sets of resonances of Fig.~\ref{fig1}(a) to merge into one 
cluster, whose individual peaks now start to become visible also at 
$T= 0.005\delta$. Notice however, that the mean-level spacing 
between the peaks is 
$\approx 0.05\, \delta$, in fact much smaller than the experimentally observed 
resonance spacing, $0.2\, \delta$.
This leads us to conclude that such a large density of resonances, 
caused by an unrealistically strong coupling to the uniform spin-wave mode, 
is {\it not} the one observed experimentally.

We now turn to the case where the p-h excitation is coupled to a nonuniform
spin-wave mode.
The exchange energy of the first nonuniform mode
is $\omega \sim \Delta (a/R)^2$, where $\Delta$ is proportional
to the exchange constant, $a$ is the lattice constant and
$R$ is the nanoparticle diameter. For a 4-nm Co nanoparticle
we find $\omega \approx 1$ meV, which is 
approximately equal to $\delta$\cite{cornell1}.
 \begin{figure}
 \includegraphics[width=3.0in,height=1.5in]{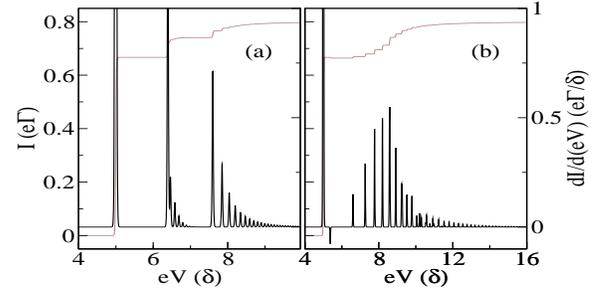}
 \caption{Current and differential conductance versus bias voltage $V$,
for $\omega = \delta$.
(a) $\gamma= 0.3 \delta$; (b) $\gamma= 0.8 \delta$
The temperature $T$ in both cases is set equal $0.005\,\delta$.} 
 \label{fig2}
 \end{figure}
In Fig.~\ref{fig2}(a) we plot the IV characteristics for the resonant case,
$\omega = \delta$, and two different values of $\gamma$. 
At small $\gamma$ we
have again two separate sets of resonances, 
which are now perfectly resolvable even at the experimental temperature. 
When $\gamma$ is increased up to $0.8$, 
the two sets of 
resonances merge into one cluster, as shown in Fig.~\ref{fig2}(b). The
number of resonances in the cluster is of the order of 15, with level
spacing $\approx 0.3\delta = 0.3(\epsilon_b-\epsilon_a) = 0.3 \omega$. 
Such a dense set of resonances with spacing
$\approx 0.2-0.5 \delta$ is one
of the characteristic features observed
experimentally in tunneling spectroscopy of magnetic
nanograins. 
The results of Fig.~\ref{fig2} do not depend on $\omega$ being exactly
equal to $\delta$ but remain valid for $\delta \ge \omega$,
although the larger $\omega$ the larger is $\gamma$ that takes to go
from Fig.~\ref{fig2}(a) to Fig.~\ref{fig2}(b).
For nanoparticles much smaller than the ones considered 
in Refs.\cite{cornell1, cornell2}, when $\delta >> \omega$,
our model predicts that the conductance spectrum should 
eventually exhibit sets of 
resonances separated by an energy $\approx \delta \propto 1/R^3$.


We finally discuss the magnetic field dependence
of the resonance spectrum. 
A crucial feature
of our analysis is based on the assumption that the two bare electronic
states
$\vert 1_a, 0\rangle$ and $\vert 1_b, 0\rangle$ have 
predominantly minority-spin character.
The fact that minority electrons dominate the tunneling transitions
had been originally
predicted in Refs.~\cite{cmc_ahm2000prl, kleff2001prb}
and was later confirmed by experiments in gated devices\cite{cornell2}.
We consider first the regime of small external fields, where the magnetic 
grain is close
to a reversal of the magnetic moment. The electronic states are coupled to the 
moment itself, and as this moves under the effect of the field, the energies of
the states will be subject 
to random fluctuations\cite{ac_cmc_ahm2002,brouwer05, usaj05}. 
Also the frequency
of the ferromagnetic mode can fluctuate strongly\cite{cmc_ac_ahm2002pap4}. 
Within our model these
fluctuations will result in a quasirandom dependence of conductance resonances
as a function of the field. At larger fields, after the reversal
has taken place, the situation is different.
The grain magnetic moment  
will point along the field and the energies of the minority states
$\vert 1_a, 0\rangle$ and $\vert 1_b, 0\rangle$ will increase linearly
with the field strength $B$, with a slope given by their effective 
$g_{a/b}$ factors, which are $\approx 2$ since spin-orbit coupling is weak.
Similarly the spin-wave energy dependence can be parameterized by 
$\omega(B) = \omega(0) + g_{\beta} \mu_{\rm B} B$\cite{cmc_ac_ahm2002pap4}.
We obtain
$\epsilon^{\pm}_k  - \epsilon_{k}^0 = {\rm const}
+ \frac{1}{2}[g_a + g_b + g_{\beta}\pm \Gamma(B)]\mu_{\rm B} B$ and
$ \epsilon^{\pm}_k - \epsilon_{k+1}^0 = {\rm const} +
\frac{1}{2}[g_a + g_b - g_{\beta}\pm \Gamma(B)]\mu_{\rm B} B$ 
for the resonance excitation
energies, where 
$\Gamma = \sqrt{(g_a - g_b - g_{\beta})^2 + 
{\rm const}/B^2}$. 
If we take 
$g_{a/b } \approx 2$ and $g_{\beta} \le 2$\cite{cmc_ac_ahm2002pap4}, we find that the
excitation energies are increasing functions of $B$.
Thus we conclude that the conductance spectrum exhibits essentially  
a monotonic linear dependence on the field, and the slopes of the 
resonance energies have the same sign.

In conclusion, we have proposed a model that describes coupled electron-magnon
excitations in a ferromagnetic metal nanoparticle.
 The conductance spectrum of single-electron tunneling exhibits a  broad 
and dense set of resonances when the coupling is of the order
of the magnon energy. The resonant peaks show Zeeman-shifts of the same
sign
as a function of the external field. Both features of the model are
in agreement with experiment. We expect that the
resonances originate from the coupling to nonuniform
spin waves; furthermore, the tunneling spectrum should break
into individual clusters for ultrasmall particles.

We would like to thank A. H. MacDonald, D. Ralph, K. Flensberg,
V. Fal'ko and G. Usaj for useful
discussions.
This  work was supported in part by the Swedish Research Council
under Grant No:621-2004-4339, the Faculty of Natural Science
of Kalmar University and by ONR under Grant N00014-02-1-0813.



\begin{thebibliography}{1}
\bibitem{foot1}
This peak in 
Figs.~\ref{fig1},~\ref{fig2} is an artifact of the model, 
where the state
$\vert 1_a,0\rangle$ is decoupled within the $n=1$ sector.
\end{thebibliography}
\end{document}